\documentclass[twocolumn,showpacs,prl,amsfonts,amsmath,amssymb,floatfix]{revtex4} 
\usepackage{color}
\usepackage{hhline}
\usepackage{mathrsfs}
\usepackage{graphicx}
\usepackage{dcolumn}
\usepackage{bm}
\usepackage{multirow}
\usepackage{booktabs}
\usepackage{afterpage}
\usepackage{amsmath}

\begin{document}

\title{Non-empirical Study of Superconductivity in Alkali-doped Fullerides Based on Density Functional Theory for Superconductors}

\author{Ryosuke Akashi$^{1}$}
\author{Ryotaro Arita$^{1,2}$}
\affiliation{$^1$Department of Applied Physics, The University of Tokyo, 
             Hongo, Bunkyo-ku, Tokyo 113-8656, Japan}
\affiliation{$^2$JST-PRESTO, Kawaguchi, Saitama 332-0012, Japan}

\date{\today}
\begin{abstract}
We apply the density functional theory for superconductors (SCDFT) based on the local-density approximation (LDA) to alkali-doped fullerides $A_{3}$C$_{60}$ with the face-centered cubic structure. We evaluate the superconducting transition temperature ($T_{\rm c}$) from first principles considering energy dependence of electron-phonon coupling, the mass renormalization, and the retardation effect. The calculated $T_{\rm c}$$=$7.5, 9.0 and 15.7 K for $A$$=$K, Rb, Cs are approximately 60 \% smaller than the experimentally observed values. Our results strongly suggest necessity to go beyond the framework of the Migdal-Eliashberg theory based on the LDA.
\end{abstract}

\maketitle

{\it -Introduction.} The doped fulleride superconductors $A_{3}$C$_{60}$ ($A$ $=$alkali metal)~\cite{Gunnarsson-review2004,Gunnarsson-review1997}, which exhibit maximum transition temperature ($T_{\rm c}$) of 40 K, have provided a fertile playground for theoretical and experimental studies. The most significant feature of the fullerides is the narrow metallic bands formed by molecular orbitals, whose energy scale compete with the vibrational frequencies and electron-electron interactions. Moreover, recent experiments revealed that the $T_{\rm c}$-volume ($V$) curve for this series shows a dome-like dependence near the superconductor-Mott insulator transition~\cite{Tanigaki-CsRb1991,Fleming-Tcrelation1991,Ganin-NatMater2008,Takabayashi-Science2009,Ganin-Nature2010}. This dependence is, similarly to the celebrated superconducting dome in cuprates, reminiscent of a crossover from weak to strong correlation in this system.

Motivated by these properties, various theoretical studies have investigated unconventional pairing mechanisms~\cite{Gunnarsson-review2004,Capone-review2009}. On the other hand, there have also been a received idea that the superconductivity in this system is explained by the conventional phonon-mediated pairing mechanism. Full s-wave gap with spin-singlet pairing~\cite{ZZhang-STS1991,Tycko-NMR1992,Degiorgi-opt1994}, C-isotope effect coefficient of $\gtrsim$0.20~\cite{Ebbesen-Nature1992,CCChen-Science1993, Fuhrer-PRL1999}, and the coherence peak in the NMR and $\mu$SR relaxation rate~\cite{Stenger-NMR1995,Kiefl-muSR1993} have been experimentally observed. In particular, in the $T_{\rm c}$-$V$ plot, the regime where $T_{\rm c}$ and $V$ positively correlate is seemingly consistent with the BCS theory; increasing $V$ results in smaller bandwidths, larger density of states (DOS) at the Fermi level, and subsequently stronger electron-phonon coupling. Hence, the applicability of the phonon mechanism is still unsettled.

The Migdal-Eliashberg (ME) theory~\cite{Migdal-Eliashberg}, which is a quite widely applicable theory of phonon-mediated superconductivity, includes the self energy with the lowest-order exchange contribution of the dressed phonons and the static screened Coulomb interaction. The ME theory described by the Kohn-Sham orbital based on the local-density approximation~\cite{Ceperley-Alder,PZ81} (KS-LDA) has enabled us to consider fine details of materials~\cite{Savrasov-Savrasov,Choi-MgB2,Margine-aniso}. Moreover, the recently developed density functional theory for superconductors (SCDFT)~\cite{GrossI,GrossII} has provided us a way to calculate $T_{\rm c}$ based on the ME theory nonempirically. The SCDFT treats effects of the interactions such as the mass-renormalization~\cite{Migdal-Eliashberg} and the retardation effect~\cite{Morel-Anderson} taking the detail of the electronic structure. The $T_{\rm c}$ calculated with the SCDFT has shown remarkably good agreement with experimentally observed $T_{\rm c}$~\cite{GrossII, Floris, Sanna}. However, its applications to molecular solids has not been reported due to its expensive computational cost. In this paper, we apply the SCDFT to fcc $A_{3}$C$_{60}$ [$A$$=$K and Rb under ambient pressure ($T_{\rm c}$$=$19 and 29 K), and Cs under the optimum pressure of 7 kbar ($T_{\rm c}$$=$35 K)] focusing on the regime where $T_{\rm c}$ and $V$ positively correlate. We calculate $T_{\rm c}$ to see if the SCDFT reproduces the absolute values and the alkali-metal dependence of the experimentally observed $T_{\rm c}$, with which we examine the applicability of the ME theory with the KS-LDA in the present system. The calculated $T_{\rm c}$ suggests that we need to consider some factors missing in the framework of the ME theory based on the KS-LDA.

{\it -Method.} In the current SCDFT~\cite{GrossI,GrossII} we solve the gap equation given by
\begin{eqnarray}
\Delta_{n{\bf k}}\!=\!-\mathcal{Z}_{n\!{\bf k}}\!\Delta_{n\!{\bf k}}
\!-\!\frac{1}{2}\!\sum_{n'\!{\bf k'}}\!\mathcal{K}_{n\!{\bf k}\!n'{\bf k}'}
\!\frac{\mathrm{tanh}[(\!\beta/2\!)\!E_{n'{\bf k'}}\!]}{E_{n'{\bf k'}}}\!\Delta_{n'\!{\bf k'}}.
\label{eq:gap-eq}
\end{eqnarray}
Here, $n$ and ${\bf k}$ denote the band index and crystal momentum, respectively, $\Delta$ is the gap function, and $\beta$ is the inverse temperature. The energy $E_{n {\bf k}}$ is defined as $E_{n {\bf k}}$=$\sqrt{\xi_{n {\bf k}}^{2}+\Delta_{n {\bf k}}^{2}}$ and $\xi_{n {\bf k}}=\epsilon_{n {\bf k}}-\mu$ is the one-electron energy measured from the chemical potential $\mu$, where $\epsilon_{n {\bf k}}$ is obtained by solving the normal Kohn-Sham equation 
$
\mathcal{H}_{\rm KS}|\varphi_{n{\bf k}}\rangle=\epsilon_{n{\bf k}}
|\varphi_{n{\bf k}}\rangle
$
with $\mathcal{H}_{\rm KS}$ and $|\varphi_{n{\bf k}}\rangle$ being the Kohn-Sham Hamiltonian and the Bloch state, respectively. The functions $\mathcal{Z}$ and $\mathcal{K}$ are the exchange-correlation kernels describing the effects of the interactions. The kernels describing the standard electron-phonon mechanism, $\mathcal{K}$$=$$\mathcal{K}^{\rm ph}$$+$$\mathcal{K}^{\rm el}$ and $\mathcal{Z}$$=$$\mathcal{Z}^{\rm ph}$, have been proposed~\cite{GrossI,GrossII}. Namely, the phonon contributions ($\mathcal{K}^{\rm ph}$ and $\mathcal{Z}^{\rm ph}$) were formulated referring to the ME theory, and the electron contribution ($\mathcal{K}^{\rm el}$) corresponds to the screened static Coulomb interaction scattering the Cooper pairs.

\begin{figure}[b]
 \begin{center}
  \includegraphics[scale=.75]{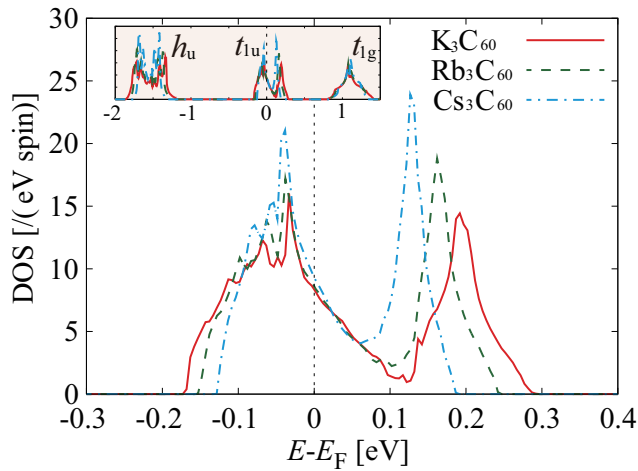}
  \caption{(Color online) DOS around the Fermi level. Inset is the view in a broader energy scale, where the characters of the bands are specified.}
  \label{fig:dos}
 \end{center}
\end{figure}

Since the fulleride superconductors involve high-frequency phonons, the electron-phonon interaction has strong dependence on both $\xi_{n{\bf k}}$ and $\xi_{n'{\bf k}'}$. In order to treat this effect, we use the $n{\bf k}$-resolved form for $\mathcal{K}^{\rm ph}$ and $\mathcal{Z}^{\rm ph}$ defined by Eqs.~(9) and (11) in Ref.~\onlinecite{GrossII}, which require the electron-phonon matrix elements $g^{\nu{\bf q}}_{n{\bf k}, n'{\bf k}'}$ and the phonon frequencies $\omega_{\nu{\bf q}}$ as inputs. For $\mathcal{K}^{\rm el}$, on the other hand, we use the form given by Eq.~(13) in Ref.~\onlinecite{Massidda}, which is based on the static random-phase approximation (RPA)~\cite{Hybertsen-Louie} and properly treats the local-field effect due to the spatial dependence of electron density. 

{\it -Result and discussion.} We calculated the band structure, phonon frequencies, electron-phonon and electron-electron interactions, and $T_{\rm c}$ for fcc $A_{3}$C$_{60}$ with $A$$=$K, Rb and Cs. All of our calculations were performed within the local-density approximation using {\it ab initio} plane-wave pseudopotential calculation codes {\sc Quantum Espresso}~\cite{Espresso,Troullier-Martins,comment-detail}. Input structural parameters~\cite{comment-struct} were determined by full relaxation and the Murnaghan equation of state~\cite{MurnaghanEOS}. For $A$$=$Cs, we considered the case of the optimum pressure of 7~kbar~\cite{Ganin-Nature2010}. Phonon frequency and electron-phonon coupling~\cite{comment-elph} were calculated by the density functional perturbation theory~\cite{Baroni-review}. The dielectric function used for $\mathcal{K}^{\rm el}$ was calculated within the static RPA. The SCDFT gap equation [Eq.~(\ref{eq:gap-eq})] was solved with the random sampling scheme given in Ref.~\onlinecite{Akashi-MNCl}, with which the sampling error in the calculated $T_{\rm c}$ was approximately 3\%.

\begin{table}[b]
\caption[t]{Calculated parameters representing the electronic structure and the electron-phonon and electron-electron interactions.}
\begin{center}
\label{tab:params}
\begin{tabular}{lccc} \toprule[2pt]
 & K$_{3}$C$_{60}$ & Rb$_{3}$C$_{60}$ & Cs$_{3}$C$_{60}$ \\
$N(0)$[/(eV spin)]&8.352 &8.609 &9.328 \\
$\lambda_{N(0)}$ &0.562 &0.570 &0.603 \\
$\lambda_{N(\xi)}$ &0.489 &0.542 &0.652 \\
$\omega_{{\rm ln},N(0)}$[K] &1071 &1054 &1052 \\
$\omega_{{\rm ln},N(\xi)}$[K] &932 &944 &940 \\
$\mathcal{Z}$ &0.350 &0.367 &0.396\\
$\mu$ &0.379 &0.370 &0.362 \\
\midrule[2pt]
\end{tabular} 
\end{center}
\end{table}

\begin{figure}[b]
 \begin{center}
  \includegraphics[scale=.70]{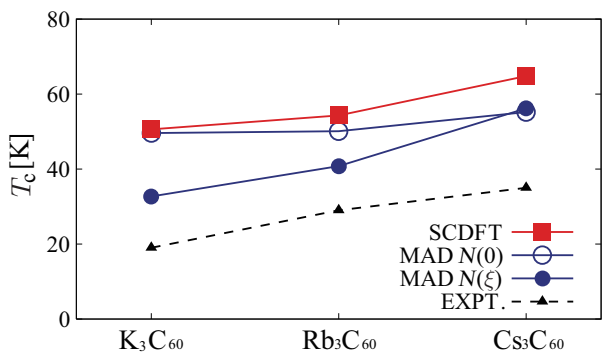}
  \caption{(Color online) Calculated $T_{\rm c}$s: Solid squares denote the values calculated using the SCDFT gap equation with only the phonon contribution ($\mathcal{K}^{\rm ph}$ and $\mathcal{Z}^{\rm ph}$), and open (solid) circles denote the values derived from the MAD formula (see text) using $\lambda_{N(0)}$ ($\lambda_{N(\xi)}$) and $\omega_{{\rm ln},N(0)}$ ($\omega_{{\rm ln},N(\xi)}$) in Table \ref{tab:params}. The triangles represent the experimentally observed values.}
  \label{fig:Tc-noel}
 \end{center}
\end{figure}
 
Let us start from the calculated DOS of the partially occupied $t_{1{\rm u}}$ bands in Fig.~\ref{fig:dos}. The general trend is consistent with the previous calculation based on the generalized gradient approximation and the experimental lattice constants~\cite{Nomura-Carbon}. As anticipated previously~\cite{Tanigaki-CsRb1991, Fleming-Tcrelation1991}, replacing lighter alkali metal elements with heavier ones (from K, Rb to Cs) leads to slightly larger DOS at the Fermi level $N(0)$(see Table \ref{tab:params}). More significantly, we also see that the bandwidth becomes narrower. The relation between these changes and the electron-phonon coupling is discussed later.

\begin{table*}[t]
\caption[t]{Experimentally observed and theoretically calculated $\Gamma$-point phonon frequencies (cm$^{-1}$). The dashes denote the splitting induced by the crystal field.}
\begin{center}
\label{tab:phonons}
\begin{tabular}{lccccccccc} \toprule[2pt]
 & \multicolumn{2}{c}{expt.} &\ &\multicolumn{3}{c}{present}&\ &\multicolumn{2}{c}{theory} \\
 & $^{a}$C$_{60}$   &$^{b}$K$_{3}$C$_{60}$ &\  & K$_{3}$C$_{60}$ &  Rb$_{3}$C$_{60}$ &  Cs$_{3}$C$_{60}$ &\ & $^{c}$K$_{3}$C$_{60}$&$^{d}$K$_{3}$C$_{60}$ \\
$H_{\rm g}$(1)& 273  &271 &\ \ \ \ \ \ \ \ \ \ \ \ \ \ &262--271&261--269&261--270 &\ \ \ \ \ \ \ \ \ \ \ \ \ \ &281&252--258\\
$H_{\rm g}$(2)& 437  &431 &\ \ \ \ \ \ \ \ \ \ \ \ \ \ &422--422&420--422&418--421 &\ \ \ \ \ \ \ \ \ \ \ \ \ \ &454&407--404\\
$H_{\rm g}$(3)& 710  &723 &\ \ \ \ \ \ \ \ \ \ \ \ \ \ &685--689&686--688&687--689 &\ \ \ \ \ \ \ \ \ \ \ \ \ \ &753&658--663\\
$H_{\rm g}$(4)& 774  &. . . &\ \ \ \ \ \ \ \ \ \ \ \ \ \ &779--779&779--780&779--783 &\ \ \ \ \ \ \ \ \ \ \ \ \ \ &785&737--740\\
$H_{\rm g}$(5)& 1099 &. . . &\ \ \ \ \ \ \ \ \ \ \ \ \ \ &1111--1116&1111--1116&1113--1120&\ \ \ \ \ \ \ \ \ \ \ \ \ \ &1091&1019--1023 \\
$H_{\rm g}$(6)& 1250 &. . . &\ \ \ \ \ \ \ \ \ \ \ \ \ \ &1268--1274&1268--1273&1271--1275&\ \ \ \ \ \ \ \ \ \ \ \ \ \ &1290&1137--1136 \\
$H_{\rm g}$(7)& 1428 &1408&\ \ \ \ \ \ \ \ \ \ \ \ \ \ &1403--1408&1402--1405&1406--1407&\ \ \ \ \ \ \ \ \ \ \ \ \ \ &1387&1349--1348 \\
$H_{\rm g}$(8)& 1575 &1547&\ \ \ \ \ \ \ \ \ \ \ \ \ \ &1532--1537&1532--1536&1532--1538&\ \ \ \ \ \ \ \ \ \ \ \ \ \ &1462&1532--1530 \\
\midrule[2pt]
\multicolumn{10}{l}{$^{a}$ Raman scattering measurement, Ref.~\onlinecite{Bethune1991}.}\\
\multicolumn{10}{l}{$^{b}$ Raman scattering measurement, Ref.~\onlinecite{Zhou1992}.}\\
\multicolumn{10}{l}{$^{c}$ \textit{Ab initio} LDA full-potential calculation based on linearized muffin-tin orbital method, Ref.~\onlinecite{Antropov1992}.}\\
\multicolumn{10}{l}{$^{d}$ \textit{Ab initio} LDA pseudopotential calculation based on the mixed basis method, Ref.~\onlinecite{Bohnen-Heid1995}.}\\
\end{tabular} 
\end{center}
\end{table*}

Table \ref{tab:phonons} summarizes our calculated frequencies of the $\Gamma$-point $H_{\rm g}$-derived modes, which are distinguished as five-fold degenerate branches with strong electron-phonon coupling. The experimentally observed and preceding theoretical frequencies are also given for comparison. The agreement between our calculation and experiments is extremely good, which illustrates that our calculation properly describes the phonon properties of the present system. The alkali-metal dependence of the frequencies is little, which is due to the intramolecular property of the modes.

We next show the $T_{\rm c}$ calculated by the SCDFT with only the phonon contributions to the gap-equation kernels ($\mathcal{K}^{\rm ph}$ and $\mathcal{Z}^{\rm ph}$). The calculated $T_{\rm c}$ (red solid square) is higher than the experimental $T_{\rm c}$, which is because of the absence of the electron contribution. These values are consistent with the recent calculation based on the Eliashberg equation~\cite{Migdal-Eliashberg} by Koretsune and Saito~\cite{Koretsune}. Interestingly, the experimentally observed alkali-metal dependence is reproduced. In order to examine the origin of this dependence, we calculated the electron-phonon coupling coefficient 
$\lambda_{N(0)}$$=$$\frac{2}{N(0)}
\sum_{{\bf k}{\bf q}nn'\nu}
\frac{|g^{\nu{\bf q}}_{n'{\bf k}+{\bf q},n{\bf k}}|^{2}}
{\omega_{\nu{\bf q}}}
\delta(\xi_{n{\bf k}})
\delta(\xi_{n'{\bf k}+{\bf q}})
$
 and the characteristic frequency 
$\omega_{{\rm ln,}N(0)}$$=$ ${\rm exp}\Big\{
\frac{2}{N(0)\lambda_{N(0)}}$ $
\sum_{{\bf k}{\bf q}nn'\nu}\!\!\!\!
\frac{|g^{\nu{\bf q}}_{n'{\bf k}\!+\!{\bf q},n{\bf k}}|^{2}}
{\omega_{\nu{\bf q}}}
\delta(\xi_{n{\bf k}})
\delta(\xi_{n'{\bf k}+{\bf q}})
{\rm ln}\omega_{\nu{\bf q}}
\Big\}
$. The calculated values are listed in Table~\ref{tab:params}. By replacing lighter alkali-metal elements with heavier ones, $\lambda_{N(0)}$ is slightly enhanced due to the increase of $N(0)$. However, when we substitute $\lambda_{N(0)}$ and $\omega_{{\rm ln,}N(0)}$ into the McMillan-Allen-Dynes (MAD) formula~\cite{MAD}, $T_{\rm c}$$=$$\frac{\omega_{\rm ln}}{1.2}{\rm exp}[-1.04(1+\lambda)/\lambda]$ (with the Coulomb pseudopotential $\mu^{\ast}$ set to 0), the dependence of the resulting $T_{\rm c}$ (blue open circle) is not as significant as that obtained from the SCDFT. Alternatively, we calculated $\lambda_{N(\xi)}$ and $\omega_{{\rm ln},N(\xi)}$ defined by the following formulae~\cite{Allen1972, Casula2011}
\begin{eqnarray}
&&\lambda_{N(\xi)}
\!=\!
\frac{2}{N(0)}
\sum_{\stackrel{{\bf k}{\bf q}}{nn'\nu}}
\frac{|g^{\nu{\bf q}}_{n'{\bf k}+{\bf q},n{\bf k}}|^{2}}
{\omega^{2}_{\nu{\bf q}}}
[f_{\beta}(\xi_{n{\bf k}})\!-\!f_{\beta}(\xi_{n'{\bf k}}\!+\!\omega_{\nu{\bf q}})]
\nonumber \\
&& \hspace{80pt}
\times \delta(\xi_{n'{\bf k}+{\bf q}}\!-\!\xi_{n{\bf k}}\!-\!\omega_{\nu{\bf q}})
\end{eqnarray}
\begin{eqnarray}
&&\omega_{{\rm ln,}N(\!\xi\!)}
\!\!=\!
{\rm exp}\Big\{\!\!
\frac{2}{N\!(\!0\!)\!\lambda_{N\!(\!\xi\!)}}\!\!\!
\sum_{\stackrel{{\bf k}{\bf q}}{nn'\nu}}\!\!\!
\frac{|g^{\nu{\bf q}}_{n'{\bf k}\!+\!{\bf q},n{\bf k}}|^{2}}
{\omega^{2}_{\nu{\bf q}}}
[f_{\beta}\!(\!\xi_{n{\bf k}}\!)\!\!-\!\!f_{\beta}(\!\xi_{n{\bf k}}\!\!+\!\omega_{\nu{\bf q}}\!)\!]
\nonumber \\
&& \hspace{80pt}
\times
\delta(\xi_{n'{\bf k}+{\bf q}}\!-\!\xi_{n{\bf k}}\!-\!\omega_{\nu{\bf q}})
{\rm ln}\omega_{\nu{\bf q}}
\Big\}
.
\end{eqnarray}
These formulae explicitly treat the energy conservation in electron-phonon scattering, and therefore include the effects of the electronic states within the phonon energy scale; since the scattering involves energy exchanges of order $\lesssim$0.2 eV, electronic states within this energy range should contribute to the pair formation. As a result, the dependence of the calculated $\lambda_{N(\xi)}$ is more noticeable than that of $\lambda_{N(0)}$, and the corresponding $T_{\rm c}$ derived from the MAD formula (blue solid circle) well reproduces the dependence of the $T_{\rm c}$ calculated by the SCDFT and the experimentally observed $T_{\rm c}$. The present analysis clarifies the significance of the electronic states within the finite energy range, not only at the Fermi level.
\begin{figure}[h]
 \begin{center}
  \includegraphics[scale=.65]{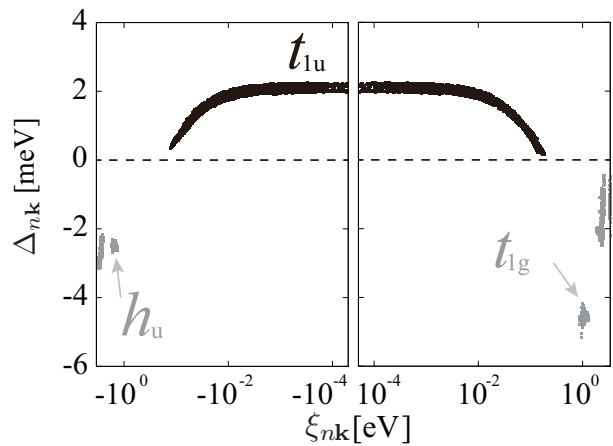}
  \caption{(Color online) Calculated gap function for Cs$_{3}$C$_{60}$ under pressure of 7 kbar with $T$$=$0.01 K. The characters of the three bands are specified.}
  \label{fig:gap}
 \end{center}
\end{figure}

\begin{figure}[b]
 \begin{center}
  \includegraphics[scale=.70]{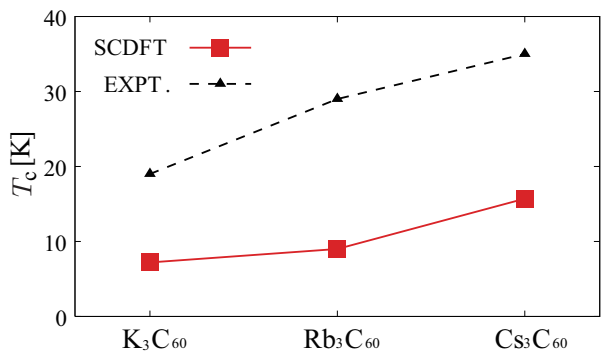}
  \caption{(Color online) Calculated $T_{\rm c}$ by solving the SCDFT gap equation with the electron contribution $\mathcal{K}^{\rm el}$ compared with the experimentally observed values.}
  \label{fig:Tc-el}
 \end{center}
\end{figure}

We also found an important aspect of the mass-renormalization factor $\mathcal{Z}$$\equiv$$\mathcal{Z}^{\rm ph}_{n{\bf k}}|_{\xi_{n{\bf k}}\rightarrow0}$ given in Table \ref{tab:params}. In usual cases, $\mathcal{Z}$ is as large as $\lambda_{N(0)}$~\cite{GrossII}, but our calculated $\mathcal{Z}$ is much smaller than $\lambda_{N(0)}$ or $\lambda_{N(\xi)}$. This is because that the $t_{1{\rm u}}$ bands are energetically isolated from other bands. The main contribution to the mass renormalization around the Fermi level generally comes from electron scattering to the states distributed within the several times of the Debye frequency. In the present case, however, the energy scale of the Debye frequency is as large as the bandwidth of the $t_{1{\rm u}}$ bands, and there is no scattering channel in the gapped region (see the inset of Fig.~\ref{fig:dos}). This weak mass renormalization results in relatively higher $T_{\rm c}$ than expected from the conventional calculations~\cite{Migdal-Eliashberg,MAD}. 

Next let us move on to the results obtained with the electron contribution ($\mathcal{K}^{\rm el}$). The strength of $\mathcal{K}^{\rm el}$ is represented by its Fermi-surface average $\mu$$=$$\frac{1}{N(0)}\sum_{n{\bf k}n'{\bf k}}\mathcal{K}^{\rm el}_{n{\bf k}n'{\bf k}}\delta(\xi_{n{\bf k}})\delta(\xi_{n'{\bf k}'})$ (See Table \ref{tab:params}). We display in Fig.~\ref{fig:gap} the gap function in $T$$=$$0.01$K for $A$$=$Cs under pressure of 7 kbar. The values of the gap function in the $t_{1{\rm u}}$ states are positive, whereas those in the highest doubly occupied $h_{\rm u}$ and the lowest unoccupied $t_{1{\rm g}}$ have negative sign. Such sign inversion of the gap function in the high energy region represents the retardation effect in the SCDFT~\cite{GrossII}. Here, the absolute values in the high energy region are quite comparable to those in the low-energy region, which signifies the strong retardation effect compared with the previously reported conventional cases~\cite{GrossII, Floris, Sanna}. This is due to large interband electron-electron Coulomb interaction.

Finally, we show the calculated $T_{\rm c}$ for $A$$=$K, Rb, and Cs in Fig.~\ref{fig:Tc-el} together with the experimentally observed $T_{\rm c}$. Thanks to the energy dependence of electron-phonon coupling, the alkali-metal dependence of the experimentally observed $T_{\rm c}$ is well reproduced. In spite of the weak mass renormalization and the significant retardation effect, the absolute values are 7.5, 9.0 and 15.7 K, which are approximately 60\% lower than the experimentally observed $T_{\rm c}$ (19, 29 and 35 K). The present underestimation is in clear contrast with the previous applications to the conventional superconductors~\cite{GrossII, Floris, Sanna}.

Since the phonon frequency scale in the present system is quite large, the theoretical $T_{\rm c}$ is sensitive to the input interactions. In fact, with $|g^{\nu{\bf q}}_{n{\bf k}, n'{\bf k}'}|^{2}$ multiplied by 1.2 (0.8), we obtain $T_{\rm c}$$=$17.5 (1.5), 20.6 (2.3), and 31.7 (5.0) K for $A$$=$K, Rb and Cs, whereas we obtain $T_{\rm c}$$=$5.8 (9.6), 7.7 (11.3), 14.7 (18.0) with $\mathcal{K}^{\rm el}$ multiplied by 1.2 (0.8). Concerning the {\it ab initio} calculation of the interactions, on the other hand, a recent paper reported that the electron-phonon interaction is enhanced by approximately 30 \% by increasing the exchange contribution in the self-consitent calculation of the wavefunctions~\cite{Janssen-Cohen}. These facts imply a possibility to fill the discrepancy between our results and experiments by considering the features neglected in the present ME theory based on the KS-LDA.

\textit{-Summary and conclusion.} Using the SCDFT, we performed non-empirical calculations of $T_{\rm c}$ in fcc $A_{3}$C$_{60}$ ($A$=K, Rb, Cs). We focused on the energy dependence of electron-phonon coupling, the weak mass renormalization and the strong retardation effect. Our calculated values of $T_{\rm c}$ were 7.5, 9.0 and 15.7~K for $A$=K, Rb and Cs (under pressure of 7 kbar), which are approximately 60\% smaller than the experimentally observed values (19, 29 and 35 K). The present results indicate a necessity to go beyond the ME theory based on the KS-LDA even for the regime where $T_{\rm c}$ and $V$ positively correlate.

{\it -Acknowledgment.} The authors thank Takashi Koretsune and Susumu Saito for fruitful discussions. This work was supported by Funding Program for World-Leading Innovative R\&D on Science and Technology (FIRST program) on ``Quantum Science on Strong Correlation", JST-PRESTO, Grants-in-Aid for Scientic Research (23340095) and the Next Generation Super Computing Project and Nanoscience Program from MEXT, Japan.

\end{document}